\documentstyle[aas2pp4]{article}
\lefthead{Zhang et al.}
\righthead{Correlation.... 4U 1820-30}

\newcommand{\etal}{{\em et al.}}

\begin{document}

\begin{flushright}
  Accepted for publication on ApJ Lett.
\end{flushright}

\title{Correlation Between Energy Spectral States and Fast Time Variability 
       and Further Evidence for the Marginally Stable Orbit in 4U 1820-30}

\author{W. Zhang, A. P. Smale\altaffilmark{1}, 
	T. E. Strohmayer, and J. H. Swank}
\affil{Laboratory for High Energy Astrophysics \\
       Goddard Space Flight Center \\
       Greenbelt, MD 20771}
\altaffiltext{1}{{\it also} Universities Space Research Association}

\begin{abstract}

We report results from a long-term monitoring data set obtained with
the Rossi X-ray Timing Explorer (RXTE) from the bright globular cluster
binary source 4U 1820-30.  During this monitoring, the source intensity
varied from a flux of $2.9\times 10^{-9} ergs\  cm^{-2} s^{-1}$ to
$8.5\times 10^{-9} ergs\  cm^{-2} s^{-1}$ in the 2-10 keV band, which
correspond to, respectively, a count rate of 1,600 cps and 4,500 cps in
the nominal 2-60 keV RXTE-PCA band. Its energy spectral state was
observed to encompass both the banana branches and the island state.
Kilohertz QPOs were observed to exist in the upper-banana and
lower-banana branches, as well as in the island state.  In particular,
we report that the kHz QPOs were present only when the soft-color
(defined as the flux ratio of the 3.9-5.0 keV band to the 1.7-3.9 keV
band) was below 0.78 without regard to the overall count rate and
without regard to whether the source is in any one of the three
branches or states.  Their centroid frequency was well correlated with
the overall count rate below 2,500 cps. When the count rate was above
2,500 cps, the QPO frequency was consistent with a constant independent
of the overall count rate. We take this as further evidence that the
inner edge of the accretion had reached the marginally stable orbit,
and therefore the constant 1060$\pm 20$ Hz is the marginally stable orbit
frequency.

\end{abstract}


\section{Introduction}

Low mass X-ray binaries are enigmatic objects. They are the brightest
X-ray sources in the sky. There is little  doubt that they harbor
fast-spinning neutron stars, yet all efforts to date of searching for
coherent pulsations in their persistent flux have resulted in negative
results (see, e.g. \cite{Vaughan_etal_1994}).  Those searches, however,
have uncovered quasi-periodic flux oscillations which are believed to
be indirectly related to the neutron star spin
(\cite{mbfm,vanderKlis_1985}).  Most recently, since their discovery with
RXTE in 4U 1728-34 (\cite{4u1728_discovery}) and Sco X-1
(\cite{scox1_discovery}), the highest frequencies observed from any
cosmic sources, the so-called kilohertz quasi-periodic oscillations,
have been observed from nearly every LMXB.  For a comprehensive review
of the literature on this subject, see van der Klis (1997).

The  bright globular cluster binary 4U 1820-30 is one of the most
studied low mass X-ray binaries because of its membership in a globular
cluster, its Type-I X-ray bursts (\cite{discovery_of_bursts}), its
variability on all time scales, its short orbital period of only
685 s, and most recently, the kilohertz quasi-periodic oscillations in
the persistent flux (\cite{4u1820_discovery}).  Since it is in 
the globular cluster NGC 6624, its distance of $6.4\pm 0.6$ kpc is
probably more accurately known to us than any other LMXB. Its short
orbital period--- the shortest known of any cosmic binary---strongly
suggests that its mass-donating star is most likely a helium dwarf with
0.06 to 0.08 solar masses (\cite{Rappaport_etal_1987}).

The characteristics of the kilohertz QPOs discovered in the persistent flux
(\cite{4u1820_discovery}) were rather similar to those of other
bursters. Their centroid frequency was quite tightly correlated with the
2-60 keV count rate.  Their fractional rms amplitudes were of the order
of 3\%. The frequency difference of the twin QPO peaks was consistent
with a constant $275\pm 8$ Hz.  Their fractional rms amplitudes
increase dramatically with energy; the 8-20 keV amplitudes are 1.8-2.9
times greater than those measured in the 1-8 keV band.

In this paper we report on data sets that sample the large count rate
and energy spectral span of this source over a calendar period of 6
months starting in February and ending in September of 1997.  In the
discussion section, we place our results in the overall context of kHz
QPO phenomenologies and discuss the most important features of our 
results.

\section{Observations}

The RXTE satellite was launched on 30 December 1995. It carries three
instruments: an array of 5 xenon gas proportional counters with a nominal energy
band of 2--60 keV, a high energy X-ray timing experiment (HEXTE) with a
bandwidth of 15--250 keV, and an all sky monitor. In this paper we
use only the data from the proportional counter array (PCA) because of
its large area and high sensitivity in detecting kHz QPOs.

The data we analyzed for this paper came from two sets of
observations.  The first set with an exposure time of 78 ks occurred
between 26 and 31 October 1996. The second set with a total exposure of
153 ks accumulated between 9 February and 10 September 1997. The
results from the first data set were reported previously in the kHz QPO
discovery paper (\cite{4u1820_discovery}). But we include those
measurements here for completeness. The second set of observations were
designed to sample the source over its super-orbital intensity
variation period of about 176 days
(\cite{discovery_of_176_day_period}). The combined data set of 231 ks
was broken up into 90 segments by the RXTE planning process, the south
Atlantic anomaly, and Earth occult of the source, with the shortest
contiguous exposure of 128 s and the longest 3,840 s.  The typical
segment lasted 3,000 s.  The data were collected in an event mode with
a 128$\mu t$ ($\mu t = 2^{-20} \mu s$) time resolution and a 64 channel
energy resolution, in addition to the standard one and two data. One
Type-I X-ray burst was observed. A timing analysis of the burst data
revealed no oscillations similar to those observed in bursts from other
sources, such as 4U 1728-34 (\cite{4u1728_discovery}).

A comparison with the long term light curve obtained with the RXTE-ASM
indicates that the second data set indeed sampled a 176 day cycle of 4U
1820-30 quite adequately. It covered both the low and high states of
the source, whereas the first data set covered the low state.  The
count rate ranges from a low of 1,500 cps to 4,500 cps in the nominal
2--60 keV of the RXTE-PCA.

\section{Data Analysis and Results}

We Fourier analyzed the data of every 128 s of observation with a time
resolution of 256$\mu t$.  Then we obtained the average power spectrum
for each of the 90 segments.  In order to display and visually inspect
each of the 90 power spectra, we rebinned each one of them by a factor
of 1024 in frequency so as to have a bin size of 8 Hz. A fit to one or
two Gaussian profiles were performed on those spectra which had
visually identifiable peaks.

The energy spectral colors were calculated using the standard two
data.  For each of the 90 segments, we obtained its energy spectrum
from the standard two data set with charged particle and diffuse cosmic
X-ray background subtracted using the standard RXTE-PCA
background estimator.

\begin{figure}
\vskip 2.5 in
 \caption{Color-color diagram of the 90 segments of data. The filled and open circles represent
	  those segments with and without obervable kHz QPO peaks in their FFT power spectra,
	  respectively. The two arrows identify the two segments whose power spectra
	  are plotted in detail in Figure~\ref{contrast}.}
 \label{color_color_diagram}
 \includegraphics{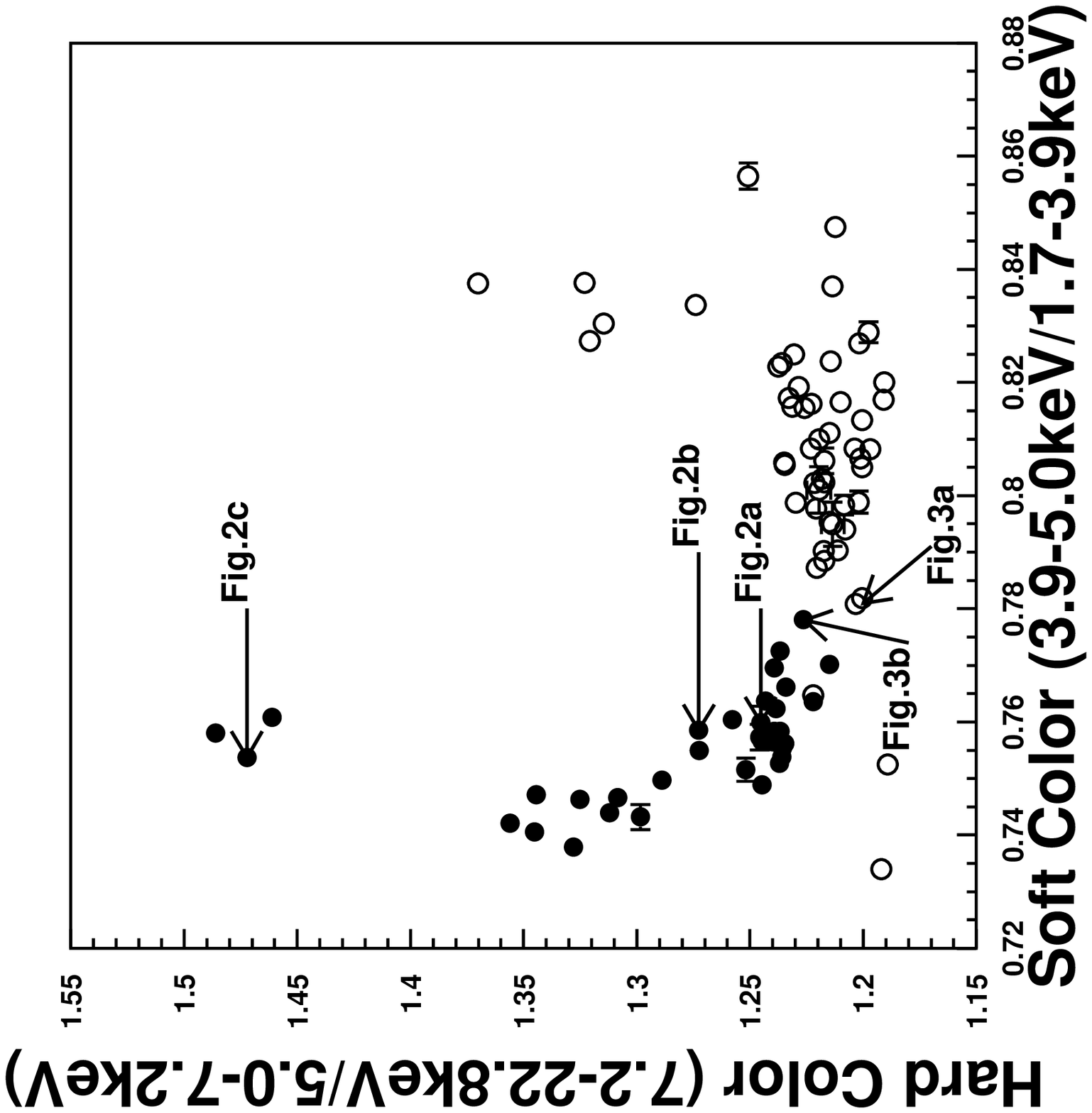}
\end{figure}

Figure~\ref{color_color_diagram} shows the X-ray color-color diagram.
Each of the 90 segments is represented by one data point. The filled
circles are those segments where kHz QPO peaks are visually identified
and numerically fitted to Gaussian profiles. The open circles are
those segments from which no QPO peaks can be identified either
visually or numerically. It is clear that with only 3 exceptions, the
open circles occupy the space to the right of soft color 0.78 and the
filled circles to the left.

\begin{figure}
\vskip 2.5 in
 \caption{Low frequency part of the power spectra of the three points as indicated in
	  Figure~\ref{color_color_diagram}. According to Hasinger and van der Klis (1989),
	  Panels a, b, and c, correspond to the upper banana branch, lower banana branch, and the 
	  island. Therefore we conclude that we have observed kHz QPOs
	  in all three branches/state of the source.}
 \label{lf_psd}
 \includegraphics{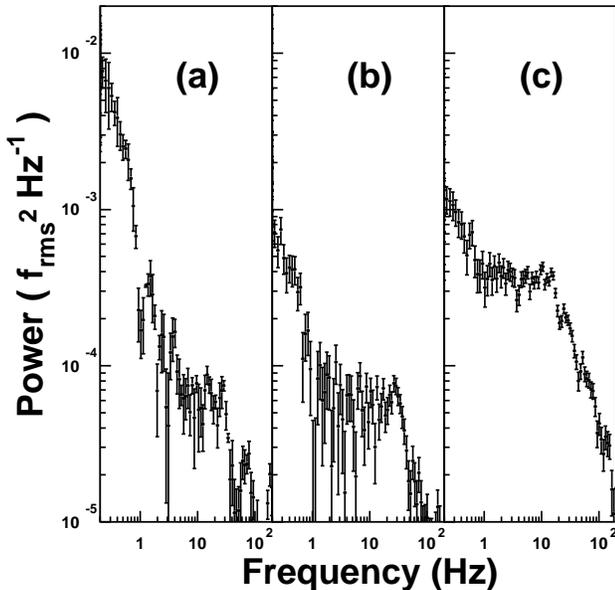}
\end{figure}

\begin{figure}
 \vskip 2.5 in
 \caption{Comparison of power spectra of two segments of data: (a) from an observation
	  on 10 September 1997 and (b) from an observation on 9 February 1997. Although the source 
	  is in very similar spectral states and has very similar count rates, 3,513 cps and
	  2,979 cps, respectively, they display rather different kHz QPO characteristics
	  as shown by this figure. }
 \label{contrast}
 \includegraphics{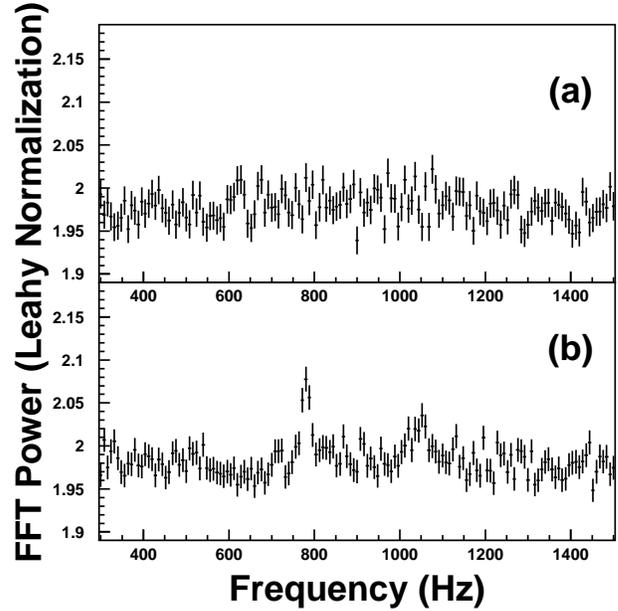}
			   
\end{figure}

In order to ascertain the nature of the energy spectral states of those
90 segments of data, we have examined their power spectra in the
frequency range of 0.02 to 100 Hz in comparison with the classification
scheme as defined by Hasinger and van der Klis (1989). We conclude that
the points in Figure~\ref{color_color_diagram} encompass both the lower
banana and the upper banana branches, as well as the island state. In
particular we note that there is not any systematic correspondence
between these branches/states and the positions in
Figure~\ref{color_color_diagram} because our data cover a large
calendar time span and it has been known that the position of the
source in the color-color diagram displays secular shifts.  In other
words, one can not read off the branch or state of the source from
Figure~\ref{color_color_diagram}.  As examples, Figures~\ref{lf_psd}
show the power spectra for the three points marked in
Figure~\ref{color_color_diagram}, which correspond to, respectively,
the upper banana branch, the lower banana branch, and the island
state.  On the one hand, we note that kHz QPOs are observed for each of
the three segments whose low frequency power spectra are shown in
Figures~\ref{lf_psd}. On the other hand, we also note that there are
other segments with very similar low frequency power spectra as shown
in the three panels of Figures~\ref{lf_psd} which do not show any kHz
QPOs. Therefore we conclude that the existence of kHz QPOs is not
directly related to whether the source is in the upper banana, lower
banana, or island state.  Rather most likely, their existence is
related to the energy spectral state as directly defined by the
hardness ratios as shown in Figure~\ref{color_color_diagram}. To
demonstrate that the separation between the open and filled circles in
Figure~\ref{color_color_diagram} is rather sharp, we show in
Figure~\ref{contrast} the FFT power spectra that correspond to the two
points as indicated by the two arrows in
Figure~\ref{color_color_diagram}.  Panel (a) is from a 10 September
1997 observation with an exposure time of 1,843 s and a count rate of
3,513 cps and (b) from a 9 February 1997 observation with an exposure
time of 2,632 s and a count rate of 2,979 cps. The measured fractional
rms amplitudes from the two peaks in panel (b) are 3.2\%$ \pm $ 0.7\%
and 2.9\%$\pm$0.7\%, respectively. Given the reasonable assumption that
any QPO in panel (a) should have the same characteristics as those in
panel (b), we set a 90\% confidence level upper limit of 1.3\% which is
typical for all other segments of data from which no kHz QPOs have been
positive detected.

Figures~\ref{f_vs_cps} and \ref{qpo_vs_cps} show the QPO
characteristics as functions of the overall source count rate in the
RXTE-PCA detectors. The centroid frequency, as shown in
Figures~\ref{f_vs_cps}  is well correlated with the count rate below
2,500 cps. Above 2,500 cps, The QPO frequencies are consistent with
constant 1060$\pm$20 Hz independent of count rate.

In the entire count rate span of 1,600 to 3,200 cps, the lower QPO and
the higher QPO centroid frequencies change in unison and maintain a
difference consistent with a constant as shown in
Figure~\ref{qpo_vs_cps}a. The fractional rms amplitude is also
correlated with the count rate over the entire count rate span as shown
in Figure~\ref{qpo_vs_cps}b.  We note here that there is ambiguity in
determining whether a QPO peak corresponds to the lower or the higher
QPO of the pair when it is the only peak observed. In
Figures~\ref{f_vs_cps} and \ref{qpo_vs_cps}, we followed the most
natural identification in that the open circles and the filled circles
form more or less smooth lines when connected.

When the count rate increases from 1,600 cps to 3,200 cps and above,
the source state transitions from the island state to the banana
branches.  At the counte rate 2,500 cps where the slope of frequency
vs. count rate changes, we do not observe any discontinuous change in
the 0.2-100 Hz parts of the power spectra. The points on the plateau in
Figure~\ref{f_vs_cps} are on the lower or upper banana branch.

\section{Discussion}

We have observed kHz QPOs from 4U 1820-30 on both the banana branches
and the island state. This is the first source from which kHz QPOs have
been observed in all branches or states (see \cite{4u1636_2qpos} and
\cite{4u1735_discovery} for discussions).  The existence of kHz QPOs in
the flux is strongly correlated with the source position on the
color-color diagram. We have examined the correlation between the
source states as defined by Hasinger and van der Klis (1989) and the
existence of kHz QPOs and found that there does not seem to be a one to
one correspondence between the two. For example, we found kHz QPOs in
all three branches/state in some segments of the data and found no kHz
QPOs in any of the three branches/state in other segments of the data.
Therefore we think that over long terms color-color daigram alone may
be a better predictor or indicator for the existence of kHz QPOs.

\begin{figure}
\vskip 2.5 in
 \caption{ QPO centroid frequencies as functions of count rate, where the 
	   filled circles represent the upper QPO and the open circles
	  the lower QPO. The two drastically different slopes are indicated by the hand-drawn
	  lines. The dates on which the data were collected are shown.}
 \label{f_vs_cps}
 \includegraphics{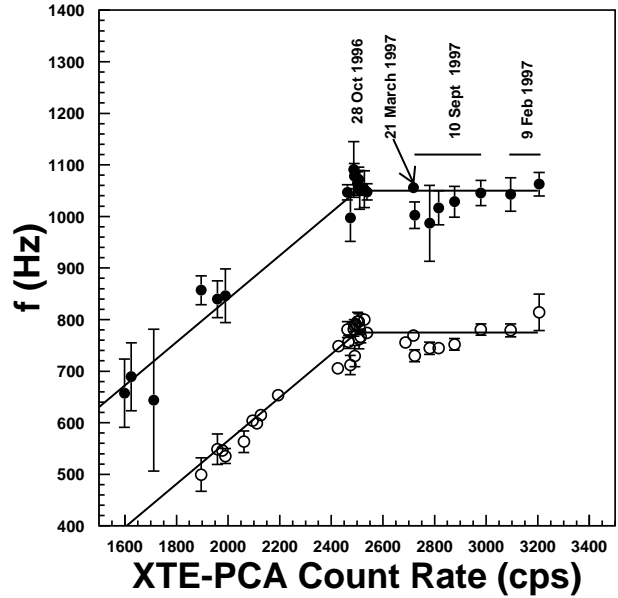}
\end{figure}

Figure~\ref{f_vs_cps} demonstrates the well-defined correlation between
the QPO centroid frequency and the overall count rate. It is similar to
nearly all other sources for which this correlation is known. Although
the data analyzed here covers a large span of calendar time, there were
no large shifts in this correlation, in contrast to the cases of 4U
0614+091 (\cite{ford_etal_4u0614}), 4U 1608-52 (\cite{yu_etal_1608}),
and Aql X-1 (\cite{aqlx1_discovery}). A very important feature of
Figure~\ref{f_vs_cps}, compared to the frequency vs. count rate
correlations of other sources, is that the QPO frequency is independent
of the count rate in the range of 2,600 to 3,200 cps.  This is a
feature that has been proposed (\cite{spm}) to strongly
support the hypothesis (\cite{kaaret_etal_1997,NS_mass_estimate_1})
that the highest observed QPO frequencies are the marginally stable
orbit frequencies near the neutron stars in the low mass X-ray
binaries.  Figure~\ref{f_vs_cps} strongly suggests that indeed we have
observed the highest QPO frequency from 4U 1820-30. If this frequency
is the marginally stable orbit frequency, then neutron star in 4U
1820-30 would have a mass in the vicinity of 2.2$M_\odot$. Given the
importance of this conclusion, more corroborating evidence from other
sources is needed.

\begin{figure}
\vskip 2.5 in
 \caption{Characteristics of the kHz QPOs plotted as functions of count rate: 
	  (a) Frequency differences where two simultaneous QPOs are observed;
	  (b) The fractional rms amplitudes, where the filled circles represent the upper QPO 
	  and the open circles the lower QPO. As in Figure~\ref{f_vs_cps}, the filled 
	  circles represent the upper QPO and the open circles the lower QPO.}
 \label{qpo_vs_cps}
 \includegraphics{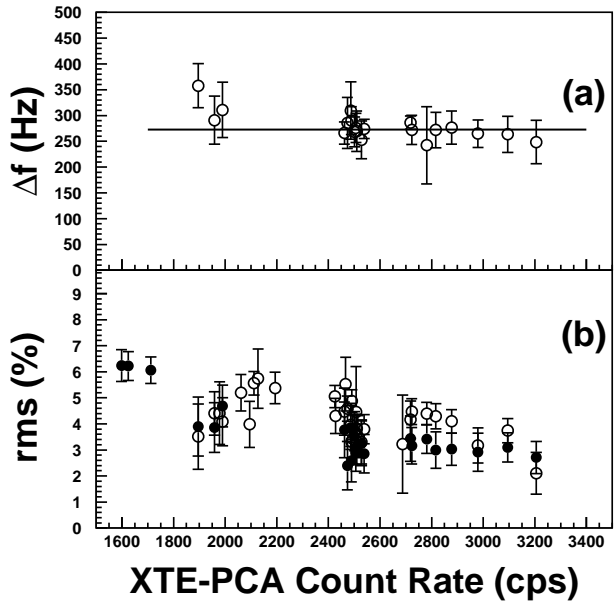}
\end{figure}

Figure~\ref{qpo_vs_cps}b tests the constancy of the frequency
difference over a much larger span of count rate than previously
possible (\cite{4u1820_discovery}). With the statistical accuracy we
can measure, the difference is a constant 275$\pm$8 Hz.  This is
similar to the cases of 4U 1728-34 and 4U 0614+091, where the
differences are consistent with being constants 363 and 327 Hz
(\cite{4u1728_discovery,4u0614_discovery}), respectively; in contrast
to the cases of Sco X-1 and 4U 1608-52 whose frequency differences have
been observed to change by over 100 Hz
(\cite{scox1_separation,4u1608_2nd_peak}).

The fractional rms amplitude as a function count rate,
Figure~\ref{qpo_vs_cps}c, indicates that, on average, when the count
rate decreases by a factor of 2 from 3,200 cps to 1,600 cps, the rms
amplitude increases by a factor of 2 from about 3\% to about 6\%. This
indicates that the flux pulsed at the QPO frequency has a constant
amplitude. It is only the un-pulsed flux that has changed by a factor
of 2. This point can be potentially constraining as it requires that
the overall X-ray luminosity of the source change by as much as a
factor of 2, but the flux pulsed at the kHz frequencies remain more or
less the same.

As the source count rate decreses, the upper frequency QPO peak
becomes broader and its amplitude larger.  The lower frequency QPO
would be sandwiched in between the low frequency noise which is below
100 Hz and the upper QPO if it were below the upper frequency by 275
Hz.  We have no positive detection of it and set a 90\% confidence
level upper limit on the fractional rms amplitude of 2\%. 

Finally we comment on the transition of the source from the banana
branch to the island state. As indicated in the original classification
paper (\cite{hk89}), our observation strongly suggests that the
transition from the banana branch to the island state is most likely a
continuous one (see the last paragraph in the last section), not an
abrupt change. As the source moves closer to the island state, i.e.,
when the count rate decreses, the upper QPO peak becomes broader and
its overall fractional rms amplitude becomes higher.  Its centroid
frequency, to the extent that we can tell from our data sets, follows
continuously along the expected curve extroplated from the banana
branch.

\acknowledgments We would like to thank the anonymous referee for
                 comments and suggestions
		 that helped us improve this paper.  W.Z. would like to
		 thank Elihu Boldt for many stimulating discussions and
		 encouragement in preparing this paper.

\pagebreak

\pagebreak

\pagebreak

\pagebreak

\pagebreak

\end{document}